# On the rotational energy distributions of reactive, non-polar species in the interstellar medium

**R.J. Glinski - E.P. Hoy - C.R. Downum**

**Abstract** A basic model for the formation of non-equilibrium rotational energy distributions is described for reactive, homo-polar diatomic molecules and ions in the interstellar medium. Kinetic models were constructed to calculate the rotational populations of $C_2^+$ under the conditions it would experience in the diffuse interstellar medium. As the non-polar ion reacts with molecular hydrogen, but not atomic hydrogen, the thermalization of a hot nascent rotational population will be arrested by chemical reaction when the $H_2$ density begins to be significant. Populations that deviate strongly from the local thermodynamic equilibrium are predicted for $C_2^+$ in environments where it may be detectable. Consequences of this are discussed and a new optical spectrum is calculated.

**Keywords** Astrochemistry - ISM: molecules - molecular processes - line: formation

R. J. Glinski, E. P. Hoy, & C. R. Downum
Department of Chemistry, Tennessee Tech University, Cookeville, TN 38501, USA
email: rglinski@tntech.edu

## 1 Introduction

We are interested in the formation and persistence of energy distributions, which deviate from the local thermodynamic equilibrium (LTE), of molecules under astrophysical conditions. In general, these deviations arise when the rates of radiative or chemical processes are comparable to the rates of thermal, Boltzmannizing collisions. In the rarified and energetic conditions of the diffuse interstellar medium (d-ISM), radiative processes are known to play a role in establishing non-LTE distributions for several non-reactive molecules. For molecular hydrogen, the phenomenon has considerable astrophysical significance and is well studied (Snow and McCall 2006). Non-LTE rotational distributions of the non-polar carbon species, $C_2$ and $C_3$, are also observed and radiative transfer models have been constructed to calculate the populations. The results of these calculations tend to agree with the observed populations (van Dishoeck and Black 1982; Roueff et al. 2002).

For reactive, non-polar species, the non-LTE effect can be even more pronounced. In the interstellar medium, a species that is reactive with $H_2$ or with H-atom can exhibit strongly non-LTE behavior—even in absence of any radiative effect. In a previous paper (Glinski et al. 1997), we examined this phenomenon in the case of the vibrational energy distributions of the reactive, non-polar species $H_2^+$, $C_2^+$, and $O_2^+$. In those cases we demonstrated that the vibrational energy distributions would maintain some of the character of the nascent distributions of the reactions in which they are formed. In this paper, we investigate the nature of non-LTE distributions that can form as steady-state populations under relatively quiescent conditions in diffuse clouds without radiative transfer. We will describe the general concept for a generic, homo-nuclear diatomic molecule. In addition, we will present approximate calculations on $C_2^+$, which is reactive with $H_2$ but not H-atom, demonstrating the necessity of including these effects in predicting its optical spectrum.

## 2 Origins of non-LTE rotational energy distributions

### 2.1 Radiative disequilibrium

Energy distributions that correspond to the LTE are formed by an equilibrium of the up and down rates between a manifold of energy levels by inelastic collisions. Non-LTE distributions can be formed when there exists a mechanism to "pump" the system in one direction (usually up) that is not in equilibrium with the rate of relaxation in the other direction. Non-LTE energy distributions formed by a radiative pumping mechanism are probably the most common. Fig. 1 shows a schematic of such a process for a generic, homo-polar, $^1\Sigma$, diatomic molecule having a stack of $J$-rotational levels in a ground state (so alternate $J$-levels are missing). Electromagnetic fields in the d-ISM can pump the molecules into an excited vibrational or electronic state, which can quickly decay back into ground rotational states. When the rate of this process is competitive with the collisional shuffling that occurs among ground-state rotational levels, a non-LTE distribution can result. For non-polar molecules, in very low number density regions, this shuffling can be relatively slow. The excitation and emission processes are both $\Delta J = \pm 2$ processes, so the net result of the pumping is a $\Delta J = \pm 4$ change in the ground-state rotational level. Because the number of states increases as $2J + 1$, the net rate is up (Le Bourlot et al. 1987). This is approximately the model that predicts the non-LTE distributions in $C_2$ and $C_3$ (van Dishoeck and Black 1982; Roueff et al. 2002).

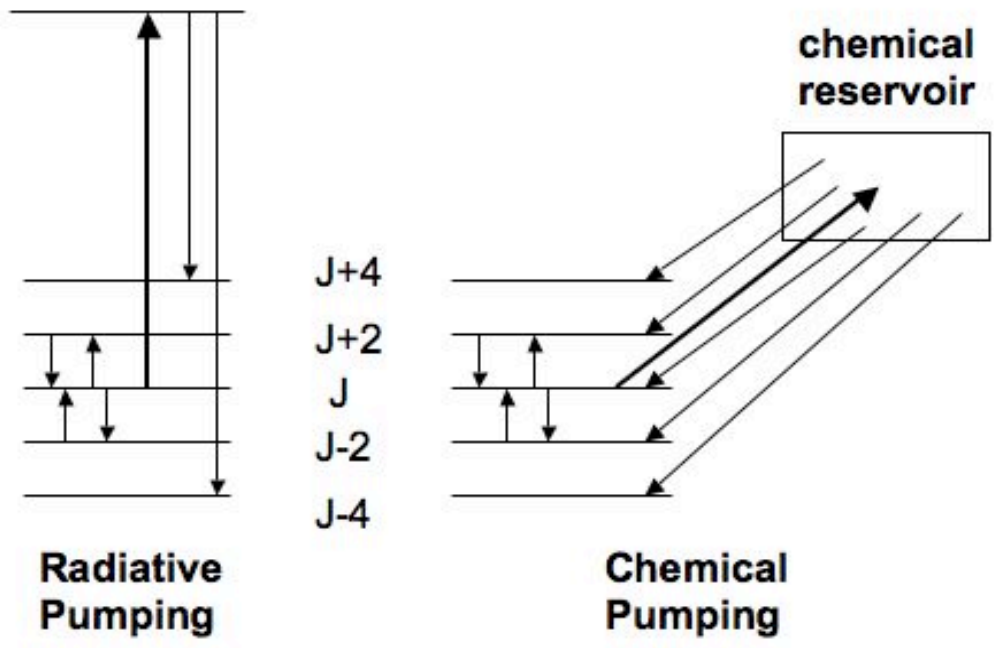

**Fig. 1** Schematics of two processes that can lead to non-LTE rotational energy distributions in a homo-polar, $^1\Sigma$ molecule. The bold arrows represent the two pumping processes: 1) absorption of a photon into an excited state or 2) chemical reaction. For a non-polar molecule, the ground-state rotational *J* levels are Boltzmannized via non-reactive collisions (short arrows).

For the $C_2$ molecule, the detailed photokinetics have been worked out in the ISM (van Dishoeck and Black 1989), with reasonable success in modeling the observations (Le Bourlot et al. 1987; Cecchi-Pestellini and Dalgarno 2002). Recently, observations have been reported of optical spectra of $C_3$ that display an intriguing array of distributions among differing lines of sight (Maier et al. 2001; Galazutdinov et al. 2002; Ádámkovics et al. 2003). Preliminary radiative transfer models have had semi-quantitative success in interpreting the general form of the $C_3$ distributions (Roueff et al. 2002). These models have necessarily included Boltzmannizing collisions with H-atom and $H_2$; but did not include chemical reaction (or photodissociation), as both the non-polar $C_2$ and $C_3$ are considered non-reactive (or slowly reactive) with the hydrogen species.

## 2.2 Reactive Disequilibrium

Chemical reaction can also lead to non-LTE distributions. This is shown schematically as the chemical pumping mechanism in Fig. 1. Every species in the ISM has a cycle of existence that includes formation and destruction processes. When the rate of destruction is comparable to the rate of rotational energy reshuffling, the result can be a non-LTE distribution. (The destruction process could include photodissociation.) The chemical process is not restricted to $\Delta J \pm 4$ jumps; therefore, chemical reaction can be more efficient than a radiative process in producing a disequilibrium. The difference is that reaction stores the species from all of the states into a chemical reservoir; the species is then reformed from the reservoir (without memory) into all of the states, with the distribution characteristic of the reforming reaction. The basic idea is that chemical reactions in space are generally exoergic and form products that carry some of the reaction energy in their internal degrees of freedom.

We have constructed a basic model to study the general effects of reaction on rotational energy distributions of a generic, homo-polar, $^1\Sigma$, diatomic molecule, having a moment of inertia approximately that of $C_2$. In the absence of chemical reactions or any radiative process, the LTE is maintained by the following physical reactions. The physical quenching reaction:

$$A_2(J) + M \rightarrow A_2(J - 2) + M \tag{1}$$

and the collisional up-stepping reaction:

$$A_2(J) + M \rightarrow A_2(J + 2) + M\,. \tag{2}$$

The collision partner M, can represent either H-atom, $H_2$, or He that are predominant in the ISM. The relative rates of reactions (1) and (2) are chosen to yield a typical LTE temperature, here 80K. If $A_2$ is a neutral molecule, the value of these rate constants are estimable as the typical gas-kinetic rates, approximately $2 \times 10^{-10}$ cm$^3$ molec$^{-1}$ s$^{-1}$. The following mechanism is an abstraction of the complex chemical network that maintains the steady-state density of the species. The species is formed from the chemical reservoir

$$\text{reservoir} \rightarrow A_2(J) \tag{3}$$

and is destroyed, or moved back into the reservoir by chemical reaction

$$A_2(J) + X \rightarrow \text{reservoir.} \tag{4}$$

The rate constant, $k_3$, represents formation of $A_2(J)$ in a $J$-distribution reflecting a nascent distribution with an excitation temperature of 2000 K, which would represent a partitioning of 8 % of a typical reaction exoergicity of 200 kJ/mol. (If X were a photon, $k_4$ would be the photodissociation rate and the modeling would yield analogous results.) The model could be run under a range of ratios of the shuffle rate, $k_1[M]$, and the reaction rate, $k_4[X]$. Results for three ratios are shown in Fig. 2.

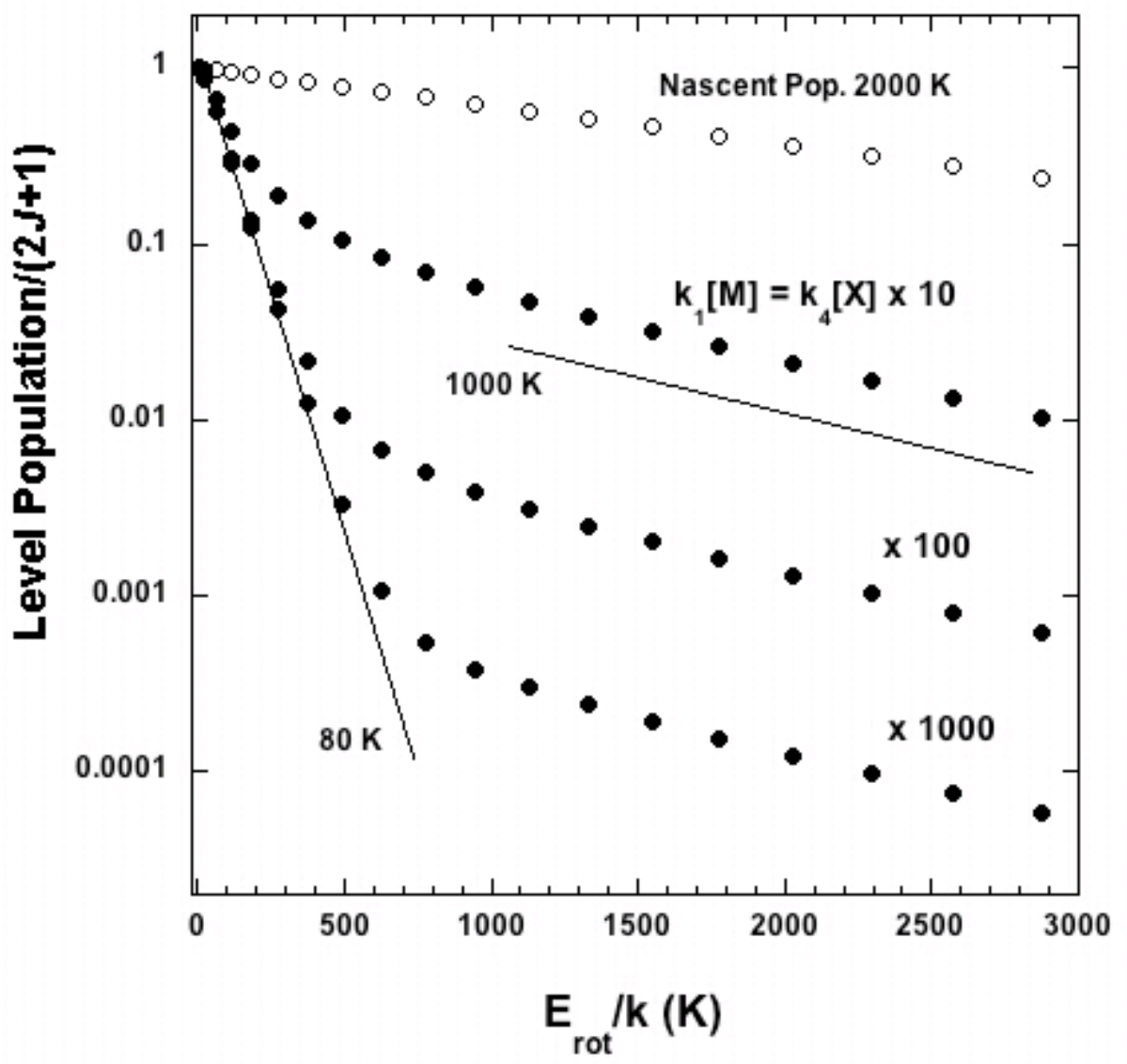

**Fig. 2** Non-LTE rotational energy distributions generated by the chemical reaction models for three different ratios of shuffle rate, $k_1[M]$, to reaction rate, $k_4[X]$. The 80 K line is the kinetic temperature used in each model. The 1000 K line represents the slope of a Boltzmann distribution at that temperature, shown for scale. The nascent rotational energy distribution used in $k_3$ is also shown.

We note the appearance and general form of two separate exponential components. The temperature of the cold component, $T_{therm}$, corresponds to the one set by the thermal collision environment (the LTE); the hot component is a product of the chemical kinetics. Several observations can be made: While $T_{hot}$ is quite high compared to $T_{therm}$; it is not as high as the temperature of the nascent distribution. It also appears that $T_{hot}$ is relatively insensitive to conditions; but generally, $T_{hot}$ decreases slightly with increasing $k_1[M]$ to $k_4[X]$ ratio. Additionally, a temperature derived by consideration of the first eight points or fewer—not recognizing that the distributions are compound—would show increasing excitation temperature with decreasing $k_1[M]$ to $k_4[X]$ ratio, but would not be a good representation of either $T_{therm}$ or $T_{hot}$. The plot above is for a neutral $A_2$ molecule. It was found that if $A_2$ were an ion and $k_1$ was for a faster Langevin rate, $k_1 \approx 1 \times 10^{-9}$ cm$^3$ molec$^{-1}$ s$^{-1}$, the model would produce similar results with respect to the $k_1[M]$ to $k_4[X]$ ratios.

For species that are un-reactive with H-atom or $H_2$, photo-processes control the formation of the non-LTE distributions. In that case, the non-LTE effects are more pronounced in more diffuse regions of an interstellar cloud where $A_v$ is lower. This is what is observed for $C_2$ where the strongest non-LTE effects are found under the conditions of lower density (van Dishoeck and Black 1982). This trend would also hold if $A_2$ were reactive with H-atom but not $H_2$. The interesting case would be when the species is reactive with $H_2$ but not H-atom.

Such is the case with $C_2^+$, where the reactivity with $H_2$ will effect the efficiency of Boltzmannization in the more dense regions of an interstellar cloud. Our interest in these studies is to examine the effects of chemical reaction on energy distributions. We will show how the differences in the expected rotational populations in regions having different ratios of molecular hydrogen to total hydrogen will provide a means of testing whether the distributions are controlled by chemistry or by radiative processes. We note that the conditions considered here, in which the non-LTE distributions are formed, are not highly energized environments, e. g., photodissociation regions. They are not conditions that are widely out of equilibrium, such as in local regions of the ISM that give rise to maser activity (Duley and Williams 1984).

## 3 Modelling for $C_2^+$ under conditions of the diffuse and translucent ISM

There is a limited number of non-polar species that react rapidly with either H-atom or $H_2$. None of the three common diatomics, $C_2$, $N_2$, or $O_2$, are reactive. The cations $C_2^+$ and $N_2^+$, however, are both reactive with $H_2$ but not with H-atom. The reactivity of $O_2^+$ with either species has not been reported (Woodall et al. 2007). As $C_2^+$ may be an important chain carrier for the production of larger carbon molecules (Smith 1992; Oka et al. 2003), we turn our attention to the rotational energy distributions in $C_2^+$ as they would appear under conditions in the ISM. We investigate how chemical reaction gives rise to strongly non-Boltzmann energy distributions in this ion. Additionally, $C_2^+$ has a known optical electronic spectrum (Maier and Rösslein 1988), which has enabled a preliminary search in diffuse interstellar clouds (Maier et al. 2001) where only an upper limit to the density could be deduced. We have recalculated the form of the optical spectrum of $C_2^+$ using our newly calculated non-LTE populations to allow future testing of our hypothesis.

We have modeled the formation, destruction, and rotational energy redistribution of $C_2^+(X^4\Sigma_g^-)$ in a bath of $H_2$ and H-atoms (and He) under two cloud conditions, one "diffuse" and one "translucent." Our model includes the dominant processes in the life-cycle of $C_2^+$ in the ISM: formation with a given nascent distribution from the chemical reservoir, inelastic collision with H-atom (and He), reaction with $H_2$, and the slow radiative magnetic-dipole cascade. We have made reasonable estimates of the rates of all the rotational energy redistribution modes for the spin-rotational states. Next, we describe the chemical and astrophysical conditions under which $C_2^+$ was studied, followed by the rotational distribution kinetic model.

### 3.1 $C_2^+$ Chemistry in the ISM

As discussed in a previous paper (Glinski et al. 1997), $C_2^+$ reacts rapidly with $H_2$, but not with H-atom, and should retain some of its nascent vibrational or rotational energy in the dense-ISM. Only a semi-quantitative estimate of the vibrational distribution was made at that time. In this paper, we take a closer look at the rotational populations that would be expected for $C_2^+$ in the diffuse-ISM. Since the species has an inherently high reactivity, it will have a low number density, obviously providing a challenge for observation.

To begin, we examine the primary formation and destruction processes for $C_2^+$ in the ISM. The two principle formation routes are:

$$C_2 + \text{photon (12.2 eV)} \rightarrow C_2^+(N, J) + \text{electron} \qquad (5)$$

and

$$C^+ + CH \rightarrow C_2^+(N, J) + H + 200\ \text{kJ mol}^{-1}; \qquad (6)$$

where *N, J* are the spin-rotation quantum numbers for the ground-state molecule, $C_2^+(X^4\Sigma_g^-)$. (Note that *J* has a different meaning here than in section 2.) Each of these processes can form $C_2^+$ with a high degree of rotational excitation. It will be the nascent distribution of rotational levels of $C_2^+$ that will be important here. For this modeling, we have chosen a nascent rotational distribution corresponding to a temperature, $T_{form}$, of 2000 K. This corresponds to about 15 kJ/mole (0.16 eV) of the excess energy from reaction (5) or (6) going into rotation, which is a low estimate based on a statistical partitioning of the reaction energy (Klippenstein and Marcus 1989). We also have investigated how the results are affected by the nature of the nascent distribution in our studies; this is discussed below.

Of equal importance is the destruction rate of the ion; the dominant loss process is:

$$C_2^+(N, J) + H_2 \rightarrow C_2H^+ + H \qquad (7)$$

The rate constant $k_7$ was made to decrease slightly with increasing rotational excitation; the form of $k_7$ is given in Table 1. Table 1 presents a summary of the rate constants for all the processes that involve $C_2^+$ together with the rates calculated under the cloud conditions given in Table 2.

### 3.2 Two cloud conditions modelled

Before running the rotational level distribution model for $C_2^+$, we needed to choose the density and conditions in the environment in which it resides. We did this by constructing a kinetic model using a contracted set of reactions of the H, C, and O chemistry. Two cloud conditions, called "diffuse" and "translucent," were chosen for the modelling differing only by the $H_2$ photorate and $A_v$. Hence, the only principle difference is that the $H_2/H_{total}$ ratio differs in each case as seen in Table 2. The chemical modeling approach was similar to that used previously (Bucher and Glinski 1999), which began with the reaction set of (Le Bourlot et al. 1993), reduced to a smaller set of relevant reactions (Ruffle et al. 2002: Rae et al. 2002). Our models used 170 reactions between 33 species. Carbon-containing molecules as large as $CH_5^+$ and $C_2H_2$ were included; the $C_3$ molecule and larger carbon chains were not. All of the rate constants were taken from the most current UMIST database (Woodall et al. 2007). Standard rates for cosmic-ray ionization and $H_2$ production on grains were used (Black and Dalgarno 1977). The models yielded steady-state concentrations of all the species in less than $10^7$ years, starting from the elements. The "translucent" condition is realistic and the calculated fractional densities are in reasonable agreement with those calculated in more comprehensive models (Kopp et al. 1996; Le Petit et al. 2004). Calculation of more accurate, absolute densities of the species was not the objective of this work.

More dense conditions, $A_v > 1$, were not considered for two reasons: First, the ratio of hydrogen as $H_2$ to the total amount of hydrogen approaches one. Since $C_2^+$ reacts with $H_2$ on every collision it seemed clear that the energy distributions would resemble the nascent ones under cloud conditions more dense than the ones considered here. It will be easy to extrapolate to a dense condition where the collisions $C_2^+$ are primarily reactive encounters with $H_2$ or infrequent Boltzmannizing collisions with trace H (and He). Second, the $C_2^+$ density will begin to decrease. The result of this would be that although the dense conditions would yield the strongest non-LTE effects in $C_2^+$, it would become harder to detect. Therefore the relatively diffuse conditions from $A_v = 0.2$ to 0.8, were chosen to correspond to where the $C_2^+$ density may be optimum and where the non-LTE effects will vary according to condition.

Conditions were designed to allow investigation of the process in the region of both the $H/H_2$ transition and the $C/C^+$ transition of diffuse clouds (Snow and McCall 2006). Although the "diffuse" condition is not entirely realistic (as $H_{total}$ was held at 200 $cm^{-3}$ and temperature to 80 K), it was chosen as the condition where H-atom is dominant and would lead to the distribution closest to the LTE. This condition is also not realistic because it would correspond to the case where $C_2^+$ should undergo some degree of radiative pumping, as in the case of $C_2$. But as there is a lack of data on spectroscopic transitions in $C_2^+$, we have not included any radiative pumping processes in this modeling. Additionally, under the "diffuse" condition, the loss rate of $C_2^+$ from dissociative recombination is faster than the rate of reaction with $H_2$; this is included in the modelling.

3.3 Rotational level distribution model

In the two cases considered here, the density of all hydrogen species are in vast excess of that of $C_2^+$. Therefore, we could reduce the formation and destruction steps of $C_2^+$ to:

$$\text{reservoir} \rightarrow C_2^+(N, J) \tag{8}$$

and

$$C_2^+(N, J) + H_2 \rightarrow \text{reservoir}. \tag{9}$$

The reservoir was treated as a pseudo-species--the density of which was given an arbitrarily large value and the rate of reaction (8) was chosen to yield the appropriate steady-state total number density of $C_2^+$ for each cloud condition studied. Two other physical reactions were included in the model; the quenching reaction:

$$C_2^+(N_u, J_u) + \text{H (or He)} \rightarrow C_2^+(N_l, J_l) + \text{H (or He)} \tag{10}$$

and the collisional up-stepping reaction:

$$C_2^+(N_l, J_l) + \text{H (or He)} \rightarrow C_2^+(N_u, J_u) + \text{H (or He)}. \tag{11}$$

The $^4\Sigma$ state of $C_2^+$ allows for a slow radiative decay, which can occur via magnetic-dipole selection rules (Hollis 1982) from the *N, J* levels according to

$$C_2^+(N, J) \rightarrow C_2^+(N, J{=}\pm 1) \tag{12a}$$

and

$$C_2^+(N, J) \rightarrow C_2^+(N\text{-}1, J{=}0 \text{ or } J{=}\pm 1). \tag{12b}$$

The line strengths of these transitions are not found in the literature for $C_2^+$, but we have developed a first-approximation method to take into account the spin-quadruplet states. Values for the line strengths, $S_{u,l}$, are available for the spin-triplet species $O_2(^3\Sigma)$ (Maréchel et al. 1997). We have calculated the rotational populations for $C_2^+$ two ways: In the first way, we calculate the relative *N, J* populations for $C_2^+$ treating it as a $^3\Sigma$ molecule, using the $S_{u,l}$ values for $O_2$. In the second way, we calculate the total relative populations of the *N*-levels, again treating it as a $^3\Sigma$ molecule, but treating the *J*-sublevels as degenerate, where the total pseudo-degeneracy of an *N*-state is $6N + 5$. In doing this, we have found that the two cases yielded similar distributions *N*-state populations. These results are shown below. We take the second result as a good approximation for the distribution of spin-quadruplet states of $C_2^+(X^4\Sigma_g^-)$. In doing so we make the following assumptions: That the $S_{u,l}$ values calculated for $O_2$ are reasonable approximations for those values for $C_2^+$, at least for the transitions between *N*-levels; and that this treatment if it could be done for $C_2^+$ using spin-quadruplet states would yield the similar distribution of relative *N*-level populations for the two cases. Also, since the $S_{u,l}$ values are not known for $C_2^+$, microwave line emissivities cannot be calculated at this time.

Hence, reactions (8) thru (12) constitute the mechanism studied. Populations of 24 *N*-rotational levels were calculated using the volume densities of H and $H_2$ corresponding to the two cloud conditions described above. The ratio of the rates of reactions (10) and (11) were chosen to correspond to the LTE temperature of 80 K in all cases. The magnetic-dipole rate for vibrational de-excitation is significant, suggesting that there will be a slow cascade from excited vibrational levels in the ground state. But as the number of ions formed in vibrationally excited state is expected to be small (Glinski et al. 1997), the cascade is expected to make only a small enhancement in the non-LTE effects; the vibrational cascade is ignored for this calculation. Rate of radiative pumping to the $B^4\Sigma_u^-$ state is estimated to be < 2E-9 $s^{-1}$, based on comparison to $C_2$ (van Dishoeck and Black 1982; van Dishoeck and Black 1989). Additionally, we have not considered formation of the $C_2^+$ directly into the $B^4\Sigma_u$ state, as this state lies 240 kJ/mole (2.5 eV) above the ground state.

## 4 Results

### 4.1 The relative rotational energy distributions

The chemical kinetic modeling yielded the steady-state rotational populations shown in Fig. 3. Both "translucent" and "diffuse" were modeled in the two ways described above. It is seen that each spin-component yields a different distribution as is seen in the $O_2$ modelling of Maréchel et al. (1997). The distribution is essentially the same as obtained when the J-sublevels are taken as degenerate for kinetic purposes. The two conditions studied differ essentially only by the $H_2/H_{total}$ ratio and it is clear that the most strongly non-LTE distribution occurs under the condition of the largest $H_2/H_{total}$. Therefore, we expect that in lines of sight where the $C_2^+$ density is optimal for detection, the rotational distribution will resemble the one at $A_v = 0.8$ and not a thermal distribution. This implies a general trend where the more pronounced non-LTE behavior will be observed in higher $E_{B\text{-}V}$ lines of sight. The appearance of rotational excitation that is less than 80 K among the lower energy states is due to the slow magnetic dipole leak.

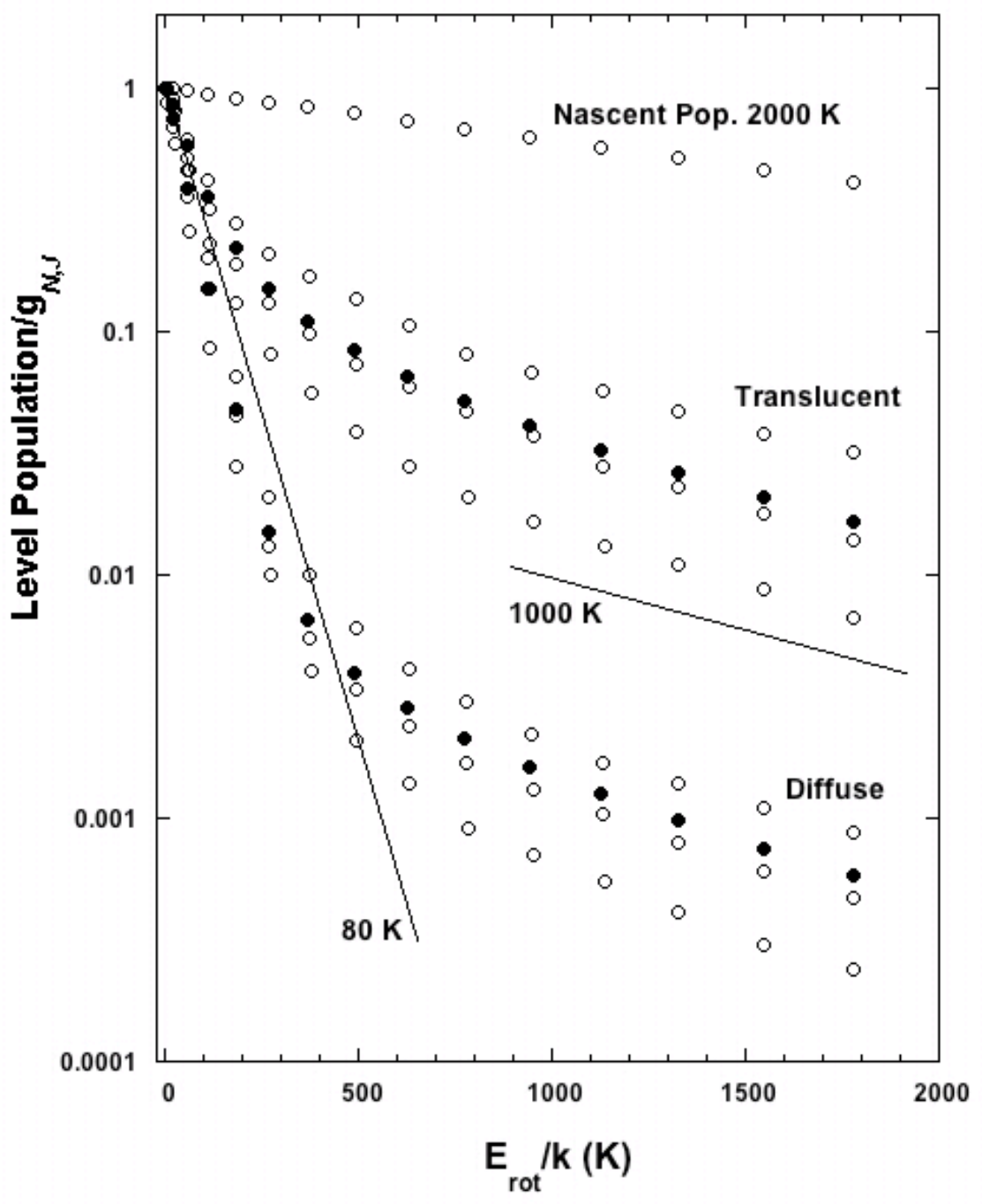

**Fig. 3** Relative rotational populations calculated under the two cloud conditions of Table 2 using two different approximations. Open circles are the results for $C_2^+$ treated as a $^3\Sigma$ species, for $J = N+1$, $N$, and $N-1$, top to bottom; closed circles are for $C_2^+$ when the $J$-substates are considered degenerate. The Boltzmann distribution temperatures have the same meaning as in Fig. 2.

## 4.2. Calculated spectra

This modeling has provided a reasonable estimate of the non-LTE rotational populations of ground state $C_2^+$ as it would likely appear in diffuse and translucent lines of sight. We have used the modelled total population of the $N$-levels to calculate the optical absorption spectrum of the system investigated by Maier and Rösslein (1988). As the spin-multiplet structure could not be resolved in that work, those authors calculated the terms using $F(N) = B_v N(N+1)$; we have done the same. We have calculated the line positions using the constants from that paper; representative spectra are shown in Fig. 4. The spectra shown in wavelength in air was calculated using the following constants: $\nu_o$(vacuum) = 19730.6 cm$^{-1}$ and B″ and B′ are 1.4281 and 1.5383 cm$^{-1}$, respectively. Both $\Delta D_e$ and $\Delta\alpha_e$ were taken to be zero. The primary differences between the two spectra are the strengths of the high-*N* lines in the R-branch to the blue of 5055 Å and the appearance of a prominent head in the P-branch at 5071.2 Å for the "translucent" condition. These features may aid in the ultimate detection of this ion.

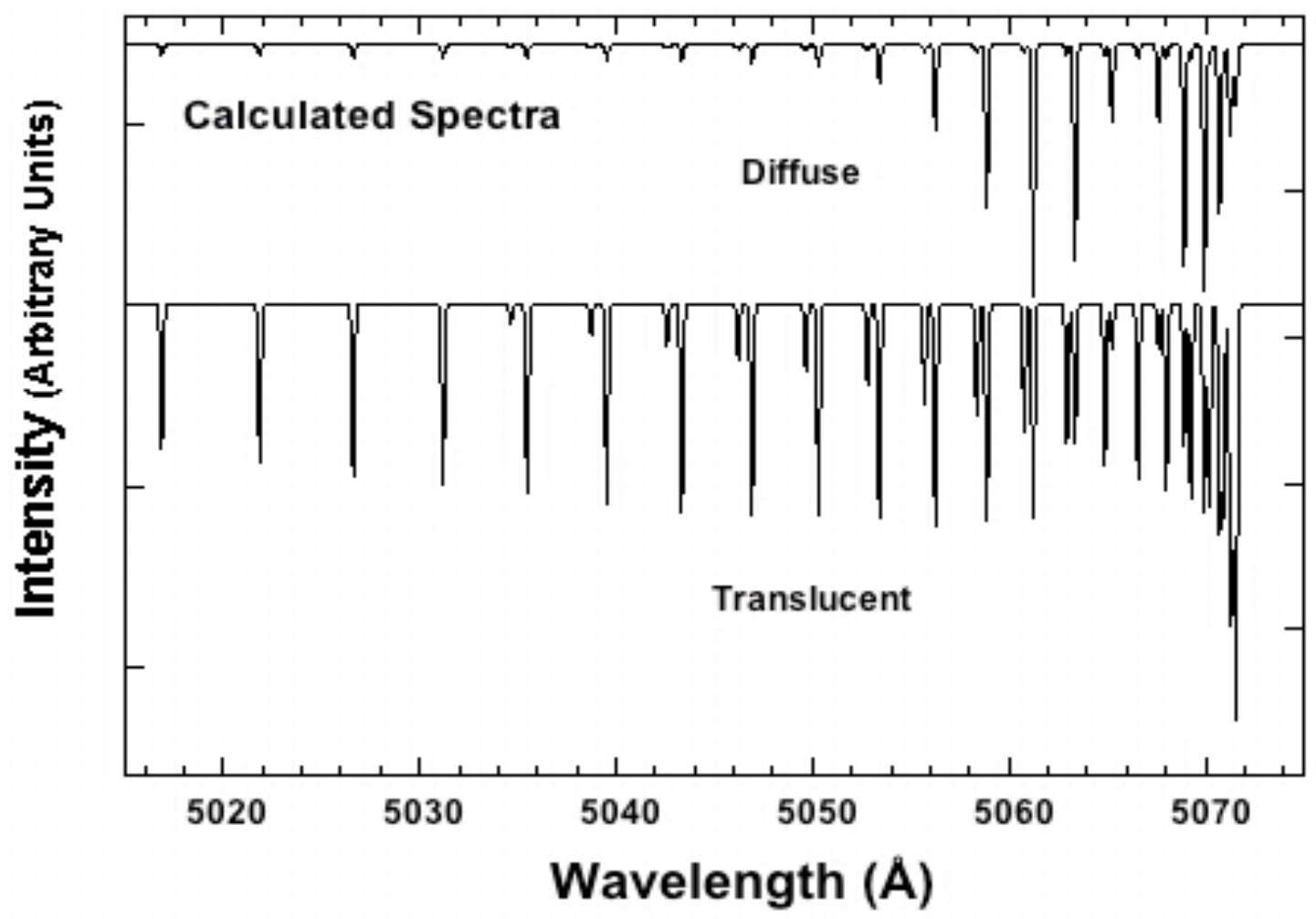

**Fig. 4** Calculated spectrum of the 0-0 band of the $C_2^+(B^4\Sigma_u^- - X^4\Sigma_g^-)$ transition the two ground state populations shown in Fig. 3. The spectral resolution (FWHM) is 0.2 Å and therefore each line contains the multiplet structure (Maier and Rösslein 1988).

We note that the spectrum for $C_2^+$ under the "diffuse" condition is almost identical to that calculated by Maier et al. (2001) for a Boltzmann distribution having temperature of 80 K. Even had those authors had the non-LTE spectrum, there are no clearly detected $C_2^+$ features in their spectrum of ζ Oph. The lack of any appearance of the band head at 5071.2 Å may require a downward revision of the upper limit of the density in that work by a factor of about five. It is clear that when $C_2^+$ inhabits an environment where the fraction of hydrogen as $H_2$ is as low as 0.25, the rotational distribution is strongly non-Boltzmann.

## 4.3 Effects of formation route

In order to investigate the influence that the formation pathway might have on the observed energy distribution, we ran the models using several different nascent distributions. The simplest exploration involved using nascent distributions having two different rotational excitation temperatures that bracketed the one chosen at 2000 K. Again we have used the model that treats the *J*-substates as degenerate. The results of that are shown in Fig. 5. Notice that the general form of the two-exponential components is maintained; but the temperature of the hot component is different. This suggests that $T_{hot}$ is related to the excitation in the nascent distribution; but this would seem difficult to interpret unambiguously. Shown also is the distribution for the “diffuse” condition, demonstrating that the stronger non-LTE effects will still be observed in the more dense lines of sight, relatively independent of $T_{form}$.

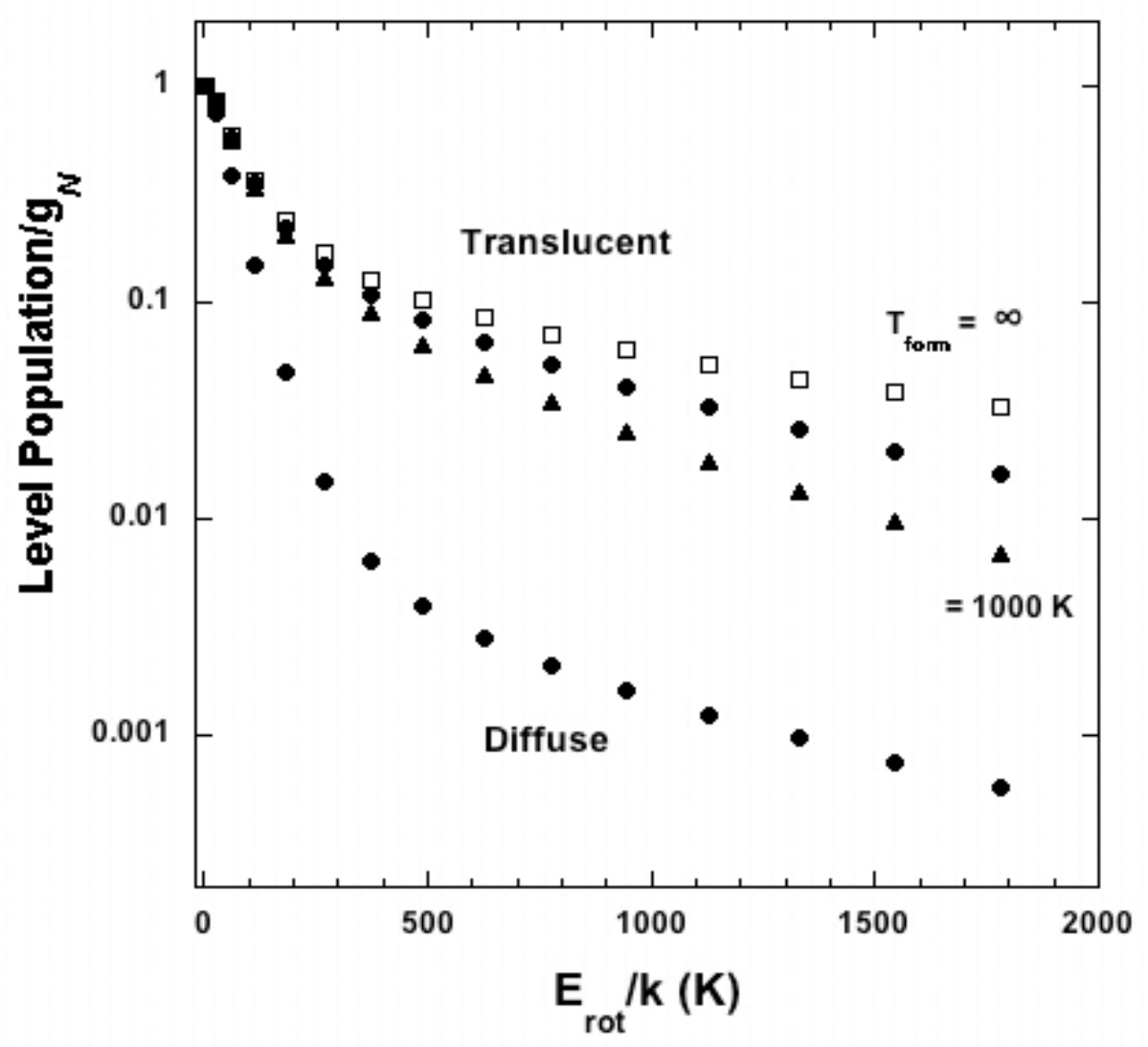

**Fig. 5** Results for the “translucent” model, when three different nascent rotational distributions are used. Solid circles are for $T_{form}$ = 2000 K. The “diffuse” model result with the original nascent distribution is included for comparison.

Since the chemical mechanism suggests that there are two possible routes that could be different dynamically, we sought to examine what the steady-state distribution would look like if the formation reaction featured a high degree of state selection in either reaction (5) or (6). To do this we increased the rates at which the $C_2^+(N)$ was reformed from the reservoir into the levels $N$ = 23, 25, 27, and 29 rates by factors of 1.3, 1.9, 1.5, and 1.3 times those at 2000 K, respectively. The increase of the rate into $N$ = 25 corresponds to an excitation temperature of about 10,000 K for that level. Additionally, in order to examine how long the spiked distribution would be manifest, we began with an initial distribution that was spiked at $N$ = 25. This was accomplished by using initial populations of the levels around $N$ = 25 that were spiked according to the ratios shown in Fig. 6. Fig. 6 presents the results of using the increased rates and spiked initial population. Steady-state was established in less than $10^{12}$ seconds; suggesting that the formation pathway will likely not be distinguishable by even careful measurement of these distributions.

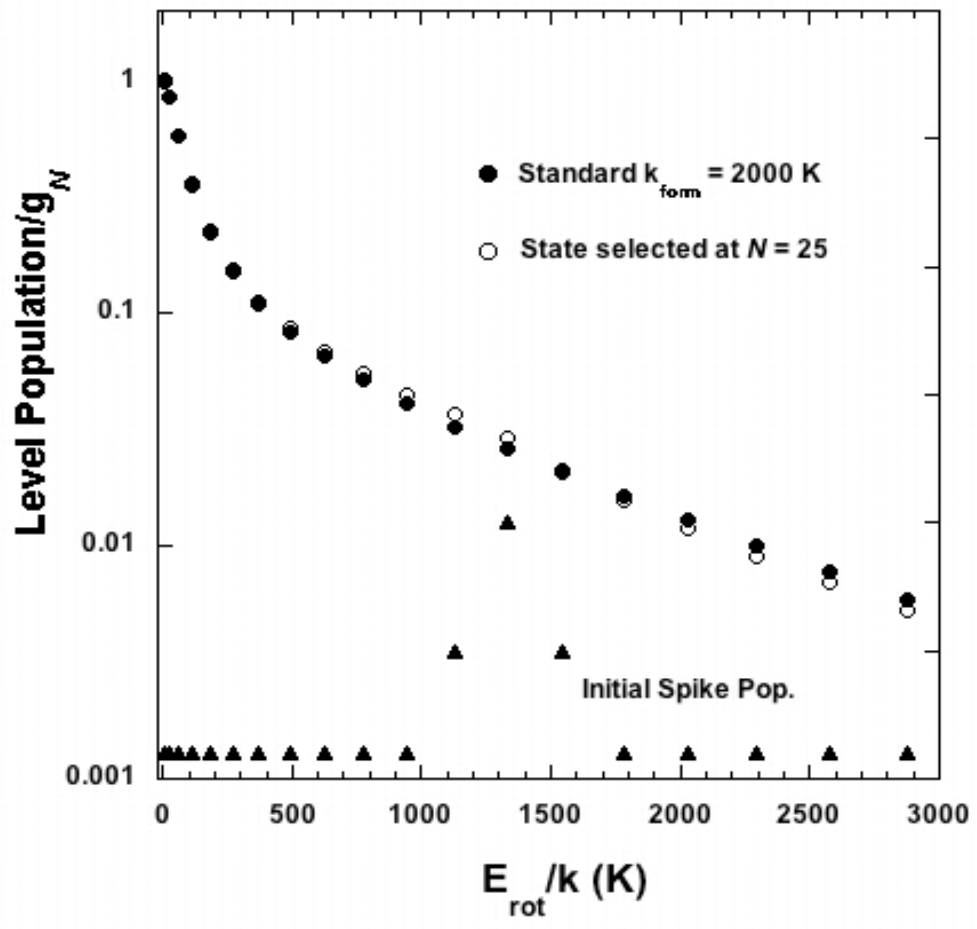

**Fig. 6** Results when $C_2^+$ is produced under the "translucent" condition with the 2000 K nascent distribution and with enhanced formation rates into several levels around $N = 25$. The populations of $C_2^+(N)$ species used as initial conditions were ten times more $N = 25$ and five times more $N = 23$ and 27 than the rest, respectively.

## 5 Discussion

Traditionally, for non-reactive species, a comparison of the total density in the environment of interest to the "critical density" determines if radiative processes need to be included in determining the rotational distribution (Tielens 2005). If the critical density, which is the value the Einstein A-coefficient for radiative rotational relaxation divided by the collision rate constant, is greater than, or on the same order as the total density, then radiative processes must be considered. Various methods have been developed and intercompared to include radiative effects in non-LTE line formation (Goldsmith and Langer 1999; van Zadelhoff et al. 2002; van der Tak et al. 2007). In such cases, the non-LTE behavior will be more pronounced under more diffuse conditions, where the interstellar radiation field is less shielded. The present work considers the case that will show the opposite behavior, where the collisions that would ordinarily establish the LTE for non-polar species are actually reactive collisions, making the traditional non-LTE analyses inapplicable.

We have previously considered the most extreme case of $H_2^+$, which is reactive on its next collision with nearly every other species (Glinski et al. 1997). In that case, no resemblance of the predicted energy distribution to the Boltzmann distribution was expected anywhere $H_2^+$ is found. The next species expected to exhibit the chemical effect, but to a lesser extent, is $C_2^+$. We have shown that the predicted non-LTE distribution will depend on the average line-of-sight density of molecular hydrogen.

We draw the following conclusions from these studies and suggest that they lead to further consideration of $C_2^+$ and the way in which non-LTE distributions are formed by chemical reaction, in general.

1. $C_2^+$ presents a case where chemical reaction competes with radiative processes and collisional relaxation to produce strongly non-Boltzmann rotational energy distributions in the ion.

2. Because $C_2^+$ reacts with $H_2$, but not H-atom, the ion should exhibit stronger non-LTE behavior under more dense conditions in the d-ISM, counter to what happens for the non-reactive $C_2$ and $C_3$.

3. The total predicted equivalent width for the optical $C_2^+$ band (Maier et al. 2001) will be spread over many more rotational levels, suggesting that non-LTE effects must be taken into account to properly describe the optical spectrum or to estimate a column density.

4. We suggest that chemistry (or photodissociation) may affect the populations of other non-polar, known or possible, interstellar species: $C_4$ and $C_5$ (Maier et al. 2004), NCCN (Petrie et al. 2003), and $HC_2C_2H^+$ (Krełowski et al. 2010). Additionally, we point out that photo-dissociation rate of $C_3$ is $6.3 \times 10^{-10}$ $s^{-1}$ and that for HCCH is $1.2 \times 10^{-9}$ $s^{-1}$, at $A_v = 1$. Therefore, if the collisional shuffle rates are on the order of $1 \times 10^{-8}$ $s^{-1}$; referring to Fig. 2, it would suggest that photo-dissociation must be included in the modelling of the rotational populations of these topical species.

5. Our modeling predicts that the chemical dynamics that produce even highly state-selected products will be smeared out in the final, observed distributions.

The following uncertainties prohibited a more comprehensive model from being devised and stronger conclusions from being made at this time: Rate constants for some of the processes, namely the rates of inelastic collisions of $C_2^+$ with H (reaction 10) are needed. The dynamics of the formation process of $C_2^+(N, J)$ would need to be understood better, in terms of the energy partitioning into rotation. Also, knowledge of the microwave line strengths, $S_{N,J}$, for $C_2^+$ and its electronic spectroscopy needs to be improved in order to better quantitatively evaluate the contribution of radiative processes, which have been approximated or ignored here.

The difficulty in observing this species is not lost on the authors. Even our simple models predict that the $C_2^+/C_2$ ratio is as low $10^{-3}$ and, therefore, the $C_2^+/C_3$ ratio will be approximately $10^{-2}$. Additionally, the optimum detectability of $C_2^+$ may be in more diffuse lines of sight, where the $E_{B-V}$ may be less than 0.5, which is at the limit of good absorption line measurements (Hobbs et al. 2008). The total column density of all $C_2^+(N, J)$ may only be $10^{10}$ $cm^{-2}$, or about $10^{-11} \times N_H$, which would make it among the most rare species detected (Snow and McCall 2006).

**Table 1** Rate constants and rates of the dominant processes for $C_2^+$ under the two conditions studied

| | $k^a$ or $A_{ul}^a$ | Rate:[b] Diffuse | Translucent |
|---|---|---|---|
| formation | | | |
| $C_2$ + photon | $4.1E\text{-}10e^{-3.5Av}$ | 6.1E-17 | 5.0E-15 |
| $C^+$ + CH | 3.8E-10 | 2.2E-16 | 4.1E-16 |
| $H^+$ + $C_2$ | 3.1E-9 | 1.9E-18 | 3.7E-16 |
| $H^+$ + $C_2H$ | 1.5E-9 | 3.6E-19 | 9.6E-17 |
| | | | |
| destruction | | | |
| $C_2^+$ + $H_2$ | 1.1E-9[c] | 4.4E-9 | 5.5E-8 |
| $C_2^+$ + photon | $1.0E\text{-}11e^{-1.7Av}$ | 9.7E-12 | 2.6E-12 |
| $C_2^+$ + electron | $3.0E\text{-}7(T/300)^{-0.5}$ | 1.2E-8 | 1.7E-9 |
| | | | |
| collisional quenching[d] | | | |
| $C_2^+(N_u, J_u)$ + H(He) → $C_2^+(N_l, J_l)$ + H(He) | 8.0E-10 | 1.5E-7 | 2.4E-7 |
| | | | |
| collisional up-stepping | | | |
| $C_2^+(N_l, J_l)$ + H(He) → $C_2^+(N_u, J_u)$ + H(He) | $8.0E\text{-}10\exp(-E_{rot}/kT_{therm})x(g_u/g_l)$ | | |
| | | | |
| magnetic dipole rates[e] | | | |
| $\Delta N = -2$ | 1E-9 -- 1E-7 | | |
| $\Delta N = 0$, $\Delta J = \pm 1$ | 7E-10 -- 9E-10 | | |

[a]Units: for bimolecular reaction, $k(cm^3\ molec^{-1}\ s^{-1})$; for photoprocesses, $A_{ul}(s^{-1})$

[b]Units: for formation, rate($molec\ cm^{-3}\ s^{-1}$); for destruction, rate($s^{-1}$)

[c]Made to decrease slowly with increasing *N, J* according to the formula:

$$1.1E\text{-}9 \times (1 + T_{rot}/300)^{-0.5}$$

[d]Chosen to be same as that for the reaction $C_2^+ + H_2$, ratioed by the Langevin rate formula

[e]Values for $O_2(X^3\Sigma^-)$ (Maréchel et al. 1997)

**Table 2** Conditions and select calculated density fractions in the two cases studied

| | Diffuse | Translucent |
|---|---|---|
| $A_v$ | 0.2 | 0.8 |
| $H_{total}$ | 200 | 400 $cm^{-3}$ |
| $H_2$ photorate[a] | 8.0E-14 | 4.0E-15 $s^{-1}$ |
| (H as $H_2$)/$H_{tot}$ | 0.02 | 0.25 |
| C/$H_{tot}$ | 4.3E-6 | 3.0E-5 |
| $C^+$/$H_{tot}$ | 9.5E-5 | 3.0E-6 |
| $C_2$/$H_{tot}$ | 7.5E-10 | 5.0E-7 |
| $C_2^+$/$H_{tot}$ | 7.5E-11 | 2.5E-10 |
| e/$H_{tot}$ | 1.1E-4 | 5.5E-6 |

Common conditions: $T_{therm}$ = 80 K, He/$H_{tot}$ = 0.10, $O_{tot}/H_{tot}$ = 2.0E-4, $C_{tot}/H_{tot}$ = 1.0E-4, Cosmic ray ionization rate = 1.2E-17 $s^{-1}$, k($H_2$)grain = 2.0E-17$n_{tot}$ $cm^3$ $molec^{-1}$ $s^{-1}$

[a]Estimated from Black and Dalgarno 1977 for these conditions.